\documentclass[9pt,twocolumn,twoside]{osajnl}

\journal{ol} 

\setboolean{shortarticle}{true}

\usepackage{caption}
\title{Deterministic generation of a perfect soliton crystal with a saturable absorber}

\author[1]{Ayata Nakashima}
\author[1,2]{Shun Fujii}
\author[1]{Riku Imamura}
\author[1]{Keigo Nagashima}
\author[1,*]{Takasumi Tanabe}

\affil[1]{Department of Electronics and Electrical Engineering, Faculty of Science and Technology, Keio University, Yokohama, Kanagawa, 223-8522, Japan}
\affil[2]{Quantum Optoelectronics Research Team, RIKEN Center for Advanced Photonics, Saitama, 351-0198, Japan}
\affil[*]{Corresponding author: takasumi@elec.keio.ac.jp}

\begin{abstract}
We numerically investigate the deterministic generation of a perfect soliton crystal (PSC) in an optical microresonator functionalized with a saturable absorber (SA). The SA allows the direct formation of a PSC from an initial, periodic Turing roll. It prevents passage through a chaotic state, which induces a stochastic nature as regards the number of generated dissipative Kerr solitons. We show that PSCs form deterministically, and the number is controlled by adjusting the input power and SA parameter. Our work provides a simple approach for obtaining a stable PSC that offers an ultra-high repetition rate and a high comb output power.
\end{abstract}

\setboolean{displaycopyright}{true}

\begin{document}

\maketitle

Studies on microresonator-based optical frequency comb devices have attracted a lot of attention owing to their compactness and ultrahigh repetition rate pulse output~\cite{doi:10.1126/science.aan8083,Tanabe_2019}. Such devices have been made possible by the development of high-$Q$ microresonators made of silica~\cite{doi:10.1063/1.4809781}, silicon nitride~\cite{Levy2010}, and crystalline~\cite{Herr2014}. The strong light confinement in a tiny space efficiently enhances light-matter interaction, allowing the generation of equally-spaced frequency components from a continuous-wave (CW) pump in an optical microresonator. Among a wide variety of cavity-enhanced nonlinear processes~\cite{Lin:16,Fujii:17,Fujii:18,Domeneguetti:21}, the four-wave mixing (FWM)-induced microresonator-based frequency comb (microcomb) and the mode-locked state of a microcomb, namely the dissipative Kerr soliton (DKS) state, have now been used in many applications, including spectroscopy~\cite{doi:10.1126/science.aah6516}, telecommunication~\cite{Marin-Palomo2017}, and distance measurement~\cite{doi:10.1126/science.aao3924}.

After pioneering studies that attempted to generate a single DKS~\cite{Jaramillo-Villegas:15,Bao:17,Xiao:20}, the generation of an $N$-free-spectral-range (FSR) perfect soliton crystal (PSC) is now gaining considerable interest since it supports a high repetition rate ($\times N$) and a high comb power ($\times N^2$) compared with a single DKS. Although we can generate multiple DKSs in the same way as that used to access a single DKS, a PSC is a unique state where the $N$-number of DKSs is evenly spaced inside a microresonator. This feature represents a significant difference from conventional multi-DKSs, where the intervals between adjacent pulses are usually unequally spaced; the periodic optical lattice trap induced by anti-mode-crossing or dispersive wave emission practically determines that the positions of the multi-DKSs are sustained~\cite{Taheri2017}. However, the irregular intervals of multi-DKSs make it challenging to adopt them for practical use, except for specific applications~\cite{Hu2020}, because unordered pulse outputs result in a complex optical spectrum. Therefore, limiting the number of DKSs was an essential task, and dedicated effort has led to the development of experimental techniques for obtaining a single DKS deterministically. In this respect, PSCs are desirable states since the ordered pulse trains can be used in the same manner as a single DKS.

Sophisticated methods have already been developed for generating PSCs, such as by using avoided mode-crossings~\cite{Cole2017,Karpov2019,Wang:18}, harmonic modulation~\cite{Lu_2020}, bichromatic pumping~\cite{Lu2021}, and nonlinear mode-coupling~\cite{PhysRevA.103.023502}. Meanwhile, recent work has described Kerr comb generation, where a single DKS is deterministically generated by injecting a pulse into a Fabry-Pérot resonator with a graphene saturable absorber (SA)~\cite{Xiao:20}, which enables passive pulse shaping as often employed in fiber-based mode-locked lasers~\cite{7268846}.

In this Letter, we demonstrate deterministic PSC generation by exploiting the SA effect. Our numerical analysis reveals that we can obtain the PSCs using a simple method that involves sweeping the pump wavelength by enabling the SA effect. The results suggest a new mechanism for the generation of PSCs, which feature high power and high repetition rate pulse sources.

\begin{figure}[t!]
\centering
{\includegraphics[width=\linewidth]{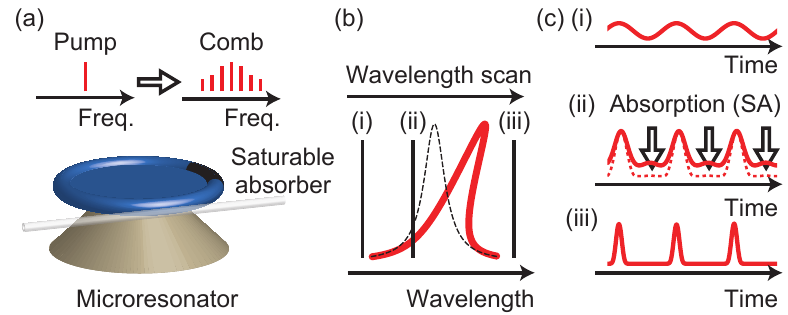}}
\caption{(a) Schematic illustration of a silica microtoroid functionalized with CNTs. (b,~c) Schematic illustrations explaining the deterministic generation of DKSs directly from a Turing pattern comb with the help of an SA. (b) is the resonance of the microresonator. (c) shows waveforms at different detunings: (i) Turing pattern, (ii) before entering the Chaos, and (iii) PSC.}
\label{fig1}
\end{figure}

Figure~\ref{fig1}(a) is a schematic illustration of our simulation. Carbon nanotubes (CNTs) are deposited on a silica ($\mathrm{SiO_2}$) toroid microresonator, which we have already employed in an experiment and that exhibits SA behavior~\cite{doi:10.1063/1.5025885}. The CW input and output lights are evanescently coupled with a tapered fiber. Figures~\ref{fig1}(b) and \ref{fig1}(c) explain our idea. When we sweep the wavelength of the input laser light, the light starts to couple with the microresonator. As a result of the modulation instability gain, the Turing rolls begin to be generated. When we further sweep the laser wavelength, the pump couples more strongly with the resonator, and the intensity of the optical field inside the resonator increases. Due to the strong nonlinear effects, the waveform starts to exhibit chaotic behavior. The DKS state is reached when the pump wavelength crosses the cavity resonance and is located on the long-wavelength detuning side. When the cavity has no SA, the number of DKS pulses is stochastic because the chaotic state appears just before the system reaches the DKS state. However, we can expect a different story if we enable the SA. After the generation of the Turing rolls, the system does not enter a chaotic state. This is due to the nonlinear absorption that suppresses the growing sub-pulses and noises, which keeps the waveform smooth. In addition, the SA effect makes the modulation depth of the Turing rolls more prominent. As a result, the system moves directly into the DKS state from the Turing rolls without exhibiting chaos. Since the modulation pattern of the initial Turing rolls is periodic, we expect periodic DKS pulses, namely PSC generation directly from the Turing rolls. Of course, we can also explain this as being because the presence of the SA helps the formation of the periodic potential trap needed to generate the PSC~\cite{Taheri2017}.

We model the evolution of the slowly varying field envelope $E(t,\tau)$ in a cavity with the Lugiato-Lefever equation (LLE). When we take the nonlinear loss of the SA into account~\cite{Suzuki:21}, it is given as,
\begin{equation}
	\begin{split}
		t_\mathrm{R}\frac{\partial }{\partial t}E=&\left(-\frac{\alpha_\mathrm{tot}}{2}-\frac{q_0}{1+|E|^2/P_\mathrm{sat}}\right)E\\
		&+\left(-i\delta_0-\frac{iL}{2}\beta_2\frac{\partial^2}{\partial \tau^2}+iL\gamma |E|^2\right)E+\sqrt{\theta}E_\mathrm{in}
\label{eq:refname1}
	\end{split}
\end{equation}
where $t_\mathrm{R}$, $\alpha_{\mathrm{tot}}$, $q_0$, $P_{\mathrm{sat}}$, $\delta_0$, $L$, $\beta_2$, $\gamma$, $\theta$, and $E_{\mathrm{in}}$, are the roundtrip time, loss of the microresonator, modulation depth of the SA, saturation power of the SA, detuning, cavity length, dispersion, nonlinear coefficient, coupling coefficient, and external pump field respectively. Here, we ignored the recovery time of the SA since CNTs show a fast recovery time~\cite{YAMASHITA2014702}. The calculation was performed using a split-step Fourier method. The parameters were as follows: $t_R = 9.05~\mathrm{ps}$ (FSR = 110 GHz), $\alpha_{\mathrm{tot}}= 2.2\times 10^{-4}$ ($Q = 5\times10^7$), $q_0 = 4.0\times10^{-3}$, $P_{\mathrm{sat}} = 4.46~\mathrm{W}$, $L = 600\pi$~\textmu m, $\beta_2 = -17.7~\mathrm{ps^2/km}$, $\gamma = 0.003~\mathrm{W^{-1}m^{-1}}$, $\theta = 5.5\times10^{-5}$, and $P_\mathrm{in} = |E_\mathrm{in}|^2 = 70~\mathrm{mW}$. These are typical experimentally obtained values~\cite{Suzuki:21}. To study the effect of the SA, we compare the calculation results obtained with and without the SAs.

\begin{figure}[t!]
\centering
{\includegraphics[width=\linewidth]{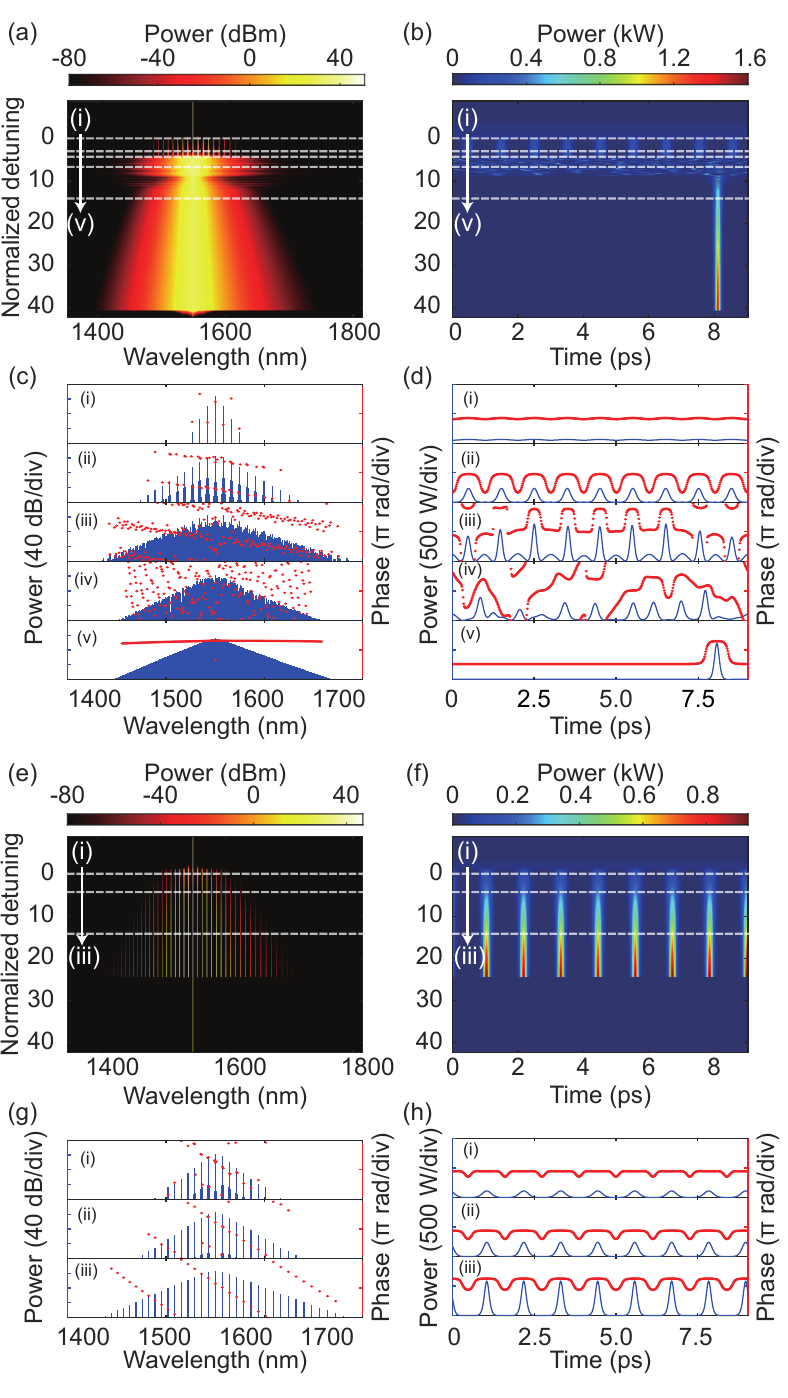}}
	\caption{(a-d) and (e-h) are results obtained without and with the SA, respectively. (a,~e) are the temporal evolution of the spectra, and (b,~f) are the waveforms when scanning the input light wavelength. (c,~d) and (g,~h) are the waveforms and spectra when the detuning is stopped at the white dashed lines in (a,~b) and (e,~f). The red dots and lines show the phases. Here, the detuning $\delta_0 = t_\mathrm{R} (\omega_0 - \omega_\mathrm{p})$ is normalized to $2\delta_0/\alpha_{\mathrm{tot}}$ ($\omega_0$ and $\omega_\mathrm{p}$ are the angular frequencies of the cavity resonance and the pump, respectively).}
\label{fig2}
\end{figure}

Figures~\ref{fig2}(a)-\ref{fig2}(d) show the evolution of the spectra and waveforms in a cavity without the SA. We observe three different regions:  Turing rolls, chaos, and DKSs. The DKS appears after chaos. Although a single DKS generation is shown in Fig.~\ref{fig2}(d), the numbers and locations are different for each calculation trial, even if the cavity and pumping parameters are unchanged.

First, we added the SA effect. We show the results in Figs.~\ref{fig2}(e)-\ref{fig2}(h). Now we observe significantly different behaviors. After the generation of the Turing rolls, the DKSs are generated directly from the Turing rolls without passing through the chaotic region. Since the initial Turing rolls have 8-equidistant peaks, we always obtain 8-equidistant DKSs. Figures~\ref{fig2}(e) and \ref{fig2}(g) show a smooth comb spectrum shape because we obtained equidistant pulses in the time domain; namely, we obtained a PSC. Neither the number nor the position changes in each calculation once we determine the cavity, the SA, and the pumping conditions. This is an interesting characteristic because the spaces between the DKSs could not be controlled if we did not use the SA. 

\begin{figure}[t!]
\centering
{\includegraphics[width=\linewidth]{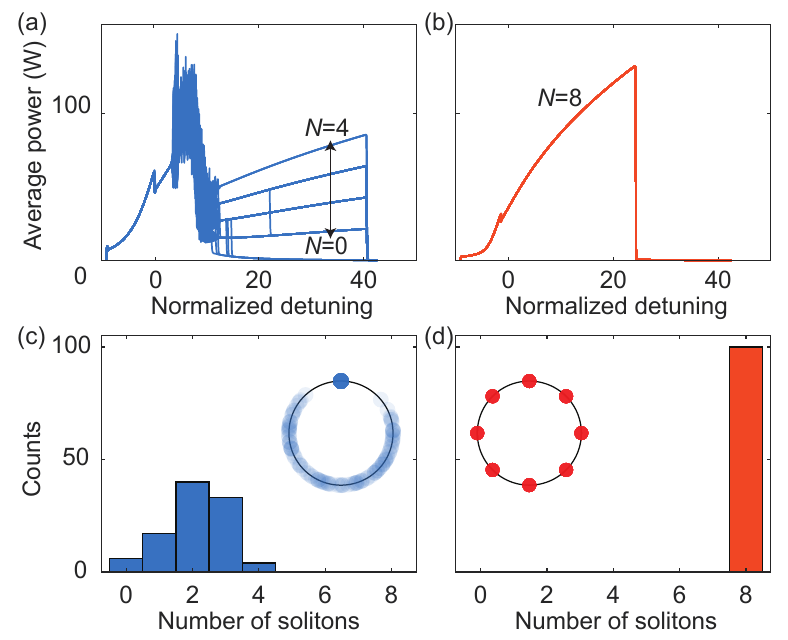}}
\caption{Results of 100 simulations. The transition of the average power inside the cavity (a) without the SA and (b) with the SA. Histogram of the number of DKSs generated in the cavity (c) without the SA and (d) with the SA. The insets in (c,~d) shows the relative positions of the generated DKSs in 100 simulations.}
\label{fig3}
\end{figure}

Figure~\ref{fig3} shows the transition of the intracavity power and the number of generated DKSs. We performed the calculation 100 times and overlay the results in Figs.~\ref{fig3}(a) and \ref{fig3}(b). Although we conducted the simulations with the same parameters, the final number of the DKS pulses $N$ is different when no SA is present, as shown in Fig.~\ref{fig3}(a). Figure~\ref{fig3}(c) is a histogram of the final number of DKS pulses. It ranges from 0 to 4, indicating that they are mostly multi-DKSs and that their number or position is deterministic. In contrast, Fig.~\ref{fig3}(b) shows the results when the SA is present. All trajectories entirely overlap, which implies that the number of DKSs is always 8 in all 100 simulations, as shown by the histogram in Fig.~\ref{fig3}(d). It is noteworthy that the intervals of the 8-DKSs are perfectly equidistant in all the calculations. Although the DKSs disappear at smaller detuning when the SA is present, which is explained by the increased loss due to the addition of the SA, the total average power is much higher in Fig.~\ref{fig3}(b) than in Fig.~\ref{fig3}(a) because of the larger $N$ in the cavity. This is another advantage of the PSC state over a single DKS state.

\begin{figure}[t!]
\centering
{\includegraphics[width=\linewidth]{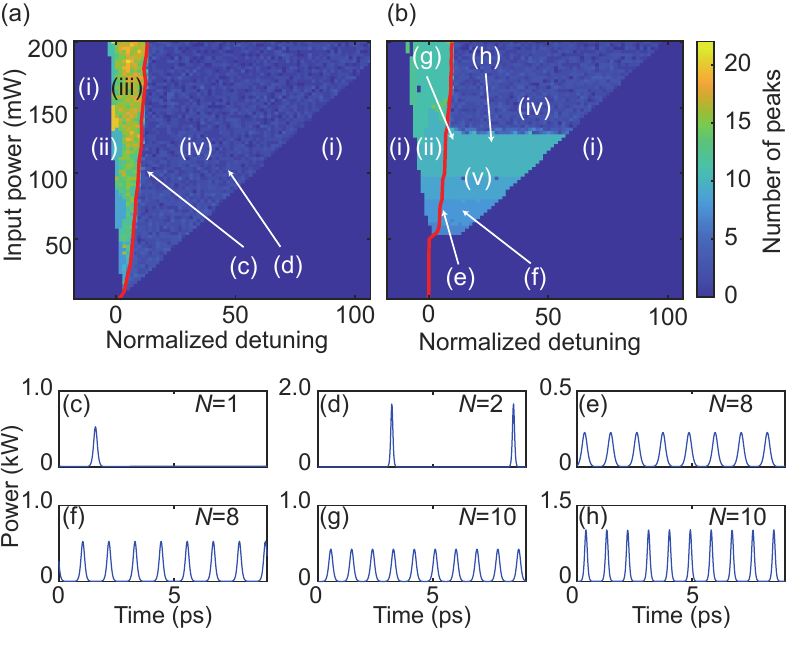}}
\caption{(a,~b) Color maps showing the number of peaks when stopping the wavelength of the pump at different detunings. (a) Without SA. (b) With SA. (c) Corresponding waveforms. The red lines in (a,~b) show the points where the effective detuning $\delta_{\mathrm{eff}}=\delta_{0}-\gamma L (|E_{\mathrm{pump}}|^2+2|E_{\mathrm{comb}}|^2)$ is zero ($|E_{\mathrm{pump}}|^2$ and $|E_{\mathrm{comb}}|^2$ are the powers at pump frequency and other frequencies, respectively).  We can distinguish the states with different color pattern regions: CW, TR (Turing rolls), Chaos, DKS, and PSC.}
\label{fig4}
\end{figure}

Next, we investigated the pump power dependence. Figures~\ref{fig4}(a) and \ref{fig4}(b) map the number of waveform peaks at different input powers $P_\mathrm{in}$ and detunings $\delta$. These maps provide information on the stability of the system. They are obtained as follows. Every point ($P_\mathrm{in}$,~$\delta$) is individually calculated with an input laser light at a constant pump power $P_\mathrm{in}$ sweeping from the shortest wavelength edge of the map until the detuning $\delta$. Then we stop the sweep and count the number of peaks. If the given parameter region exhibits chaotic or stochastic behavior, the number of pulse peaks $N$ will vary for every calculation. This results in a mosaic pattern because the colors of adjacent points will be different. Note that the number of peaks $N$ is equal to the number of DKSs in the DKS state, whereas $N$ is relatively large in the chaotic state because it is the number of peaks in a random waveform. On the other hand, if $N$ is deterministic, the region will have a monotone color. When no SA is applied, the number of DKSs is stochastic, as shown in (i) and (ii) in Fig.~\ref{fig4}(c) , which is also evident from the mosaic blue pattern in Fig.~\ref{fig4}(a). In contrast, when the SA is present, we observe a plateau color map region when the input power is located between 50 and 130~mW, as shown in Fig.~\ref{fig4}(b). In this monotone region, PSCs are generated directly from the Turing rolls. When the input power exceeds 130~mW, the number of DKS pulses is random. This is because the SA effect is not sufficiently strong to prevent the system from exhibiting a chaotic state.

\begin{figure}[t!]
\centering
{\includegraphics[width=\linewidth]{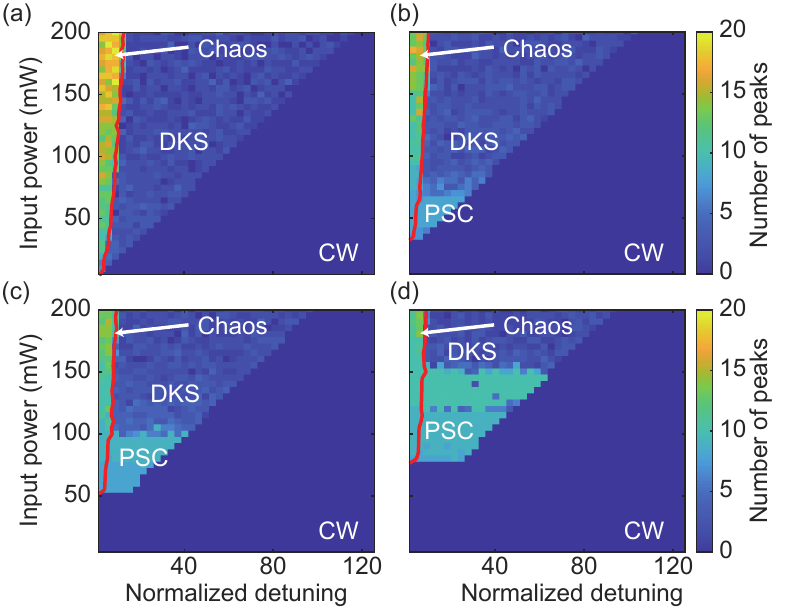}}
	\caption{Stability charts for different modulation depths. (a) $q_0 = 0$. (b) $q_0 = 3\times10^{-3}$. (c) $q_0 = 4\times10^{-3}$. (d) $q_0 = 5\times10^{-3}$.}
\label{fig5}
\end{figure}

Finally, we performed simulations with a different modulation depth $q_0$ to investigate in further detail the effect of the SA on DKS generation. Experimentally, different $q_0$ values can be achieved by varying the concentration of CNTs deposited on the resonator~\cite{Sobon2017}. Figure~\ref{fig5} shows the results for different $q_0$ values. Figure~\ref{fig5}(a) shows the results for $q_0 = 0$. When we increase the SA effect, we observe deterministic PSC generation even at a large input power (Figs.~\ref{fig5}(b)-\ref{fig5}(d)). This agrees with our understanding that the SA prevents the system from exhibiting chaos. It is noted that the lower boundary of the PSC generation region shifts toward a higher input power as we increase the $q_0$, because the overall loss also increases. In Fig.~\ref{fig5}(d), we can see a different trend. When the input power is larger than 135~mW, we observe periodic changes in the amplitude of the DKS pulses (See Fig.~S1 in Supplement~1). This breathing behaivior is beyond the scope of this paper and further work is needed to investigate this phenomenon.

Importantly, the number of DKS pulses is determined by the initially generated Turing rolls. Thus, we can control the number of pulses by designing the cavity dispersion or by choosing a different transverse mode for the microtoroids~\cite{Kato:17} since primary sidebands appear from the modes at which parametric gain reaches its maximum value~\cite{Herr2012,Fujii:18}. Figure~\ref{fig4}(b) provides evidence for the trend that a higher pump power generates Turing rolls with a shorter period, thereby increasing the number of pulses in a PSC. The details at the boundary region where the period of Turing rolls changes are found in Fig.~S2 in Supplement~1.

In this paper, we studied the generation of DKSs in a microresonator with an SA. By solving the LLE taking the SA effect into account, we found that the presence of SA significantly impacts the generation of DKSs. With the resonator without the SA, a few DKSs are stochastically generated, and their temporal intervals are random. In contrast, when we functionalized the cavity so that it had an SA, we obtained a PSC. It is noteworthy that we obtained identical results in every calculation. This is because the DKSs are generated directly from the Turing rolls, and the intracavity power does not pass through the chaotic region. This in turn implies that the number of DKSs is controlled by changing the period of the Turing rolls, for example, by appropriately designing the dispersion of the resonator~\cite{2020Nanop...9..497F}. We also investigated the principle behind the PSC generation and confirmed that the SA creates a parameter region where we can obtain stable PSCs. The parameter region is dependent on the strength of the SA effect. We believe that our work provides a new way to realize lasers with a high repetition rate and high power.

We thank Prof. A. Uchida from Saitama University for valuable discussions on the nonlinear dynamics of the cavity system. S.Fujii is supported by the RIKEN Special Postdoctoral Program.

\begin{backmatter}
\bmsection{Funding} This work was supported by JSPS KAKENHI (JP19H00873) and Amada Foundation. Part of this work is based on results obtained from a project, JPNP16007, commissioned by the New Energy and Industrial Technology Development Organization (NEDO).
\bmsection{Disclosures} The authors declare no conflicts of interest.
\bmsection{Data availability} Data underlying the results presented in this paper are not publicly available at this time but may be obtained from the authors upon reasonable request.
\bmsection{Supplemental document}
See Supplement 1 for supporting content. 
\end{backmatter}






\end{document}